\documentclass[sigconf]{acmart}

\renewcommand\footnotetextcopyrightpermission[1]{}
\acmConference[CHIWORK '26 Workshop]{%
  CHIWORK '26 Workshop: Interrogating GenAI Augmentation for CHIworkers}{%
  June 22, 2026}{Linz, Austria}
\acmYear{2026}

\newcommand{\workshopNote}{%
  \begin{quote}
    \small\itshape
    This position paper was presented at the CHIWORK '26 Workshop
    \emph{Interrogating GenAI Augmentation for CHIworkers:
    Strategies for Professional Autonomy and Accountability}
    (June 22, 2026, Linz, Austria). Workshop proposal:
    \cite{sandhaus2026interrogating}.
  \end{quote}%
}

\title{Designed by Journalists, but Is It for Readers? Rethinking AI Disclosures and Transparency in News}

\author{Pooja Prajod}
\email{Pooja.Prajod@cwi.nl}
\affiliation{%
  \institution{Centrum Wiskunde \& Informatica}
  \city{Amsterdam}
  \country{the Netherlands}
}

\begin{abstract}
As newsrooms integrate generative AI, journalists face a disclosure challenge: how to communicate AI involvement in ways that maintain reader trust. Current practice offers two approaches: brief one-line labels or detailed disclosures specifying human oversight, editorial accountability, and error reporting mechanisms. Neither achieves journalists' goal of building trust through transparency. An existing controlled experiment with 34 news readers show that detailed disclosures trigger a \textit{transparency dilemma}, reducing trust rather than increasing it, and risk introducing dark patterns that readers scroll past with the illusion of transparency. One-line disclosures avoid this effect but can create an information gap, prompting readers to expend cognitive effort searching for signs of AI involvement that the disclosure indicates but does not explain. Yet readers are not rejecting transparency, they proposed disclosure designs centered on user agency: detail-on-demand interactions, proportional AI-ratio visualizations, outlet-level signals, and explicit ``no AI'' labels. I argue that this disconnect between what practitioners believe is responsible disclosure and what users actually need is a design problem for the HCI community.
\end{abstract}

\begin{document}
\maketitle

%% Print the workshop identification note (required)
\workshopNote

%% ─────────────────────────────────────────────────────────────
\section{Introduction}
%% ─────────────────────────────────────────────────────────────

Newsrooms are becoming GenAI workplaces. Journalists increasingly use AI tools for drafting, summarizing, and translating content~\cite{beckett2023generating}. A central accountability question in this context is disclosure: how should these AI-augmented workflows be communicated to the people who consume the news? Unlike many HCI work contexts where the end user may be internal, journalism faces its users directly, i.e., readers' trust has both professional and economical implications.
 
My position is that AI disclosure in news is being treated as an ethics compliance exercise, and it is failing because practitioners are not designing for readers. The EU AI Act does not require disclosure for AI-assisted text with human editorial oversight~\cite{caisa2026}, yet some journalists/news outlets disclose AI-use voluntarily, driven by professional transparency principles. Current guidelines recommend additional information to these disclosures for better transparency, for example, affirming human oversight, identifying responsible editors, and offering error-reporting channels~\cite{BBC, NordicAI, valenzuela2026effects}. The assumption is that more transparency builds trust. A controlled experiment with 34 participants~\cite{prajod2026full} show an opposite effect. Detailed disclosures triggered a \emph{transparency dilemma}: additional context reduced rather than increased trust. One-line disclosures avoided this effect, but a brief label like ``this article was edited using AI'' can trigger an information gap, prompting readers to spend additional cognitive effort scanning the text for signs of AI involvement~\cite{prajod2026towards}. Neither approach achieves what journalists intend.
 
But readers are not asking for no disclosures. In the same study, participants proposed fundamentally different \emph{disclosure designs}, ones that give them agency over when and how much transparency they engage with. These reader-generated concepts, which I visualize in this paper, reframe disclosure from a compliance obligation into a design challenge. These visualizations suggest that journalists need a different disclosure toolkit, one designed around how readers actually engage with transparency, not around what practitioners assume is helpful.

%% ─────────────────────────────────────────────────────────────
\section{Background and Motivation}
%% ─────────────────────────────────────────────────────────────
 
The integration of GenAI into professional workflows raises a recurring question: how should practitioners communicate AI involvement to the people affected by their outputs? In scientific dissemination, this surfaces as AI use statements in papers; in social media, as disclosure icons; in journalism, as labels or statements on AI-assisted articles. Journalism is a special case because the stakes are visible and measurable. Readers' trust directly determines engagement, subscription decisions, and has economic implications for the outlet~\cite{nanz2025ai, prajod2026full}.
 
The EU AI Act adds a regulatory dimension. While the Act mandates transparency for AI-generated image, audio, and video contents, it exempts text content that has undergone human editorial review~\cite{caisa2026}. This creates a gap: the regulation assumes human oversight resolves the transparency concern, but users may not share that assumption. A growing body of research shows that there is a mismatch between the regulations and expectations of the audience~\cite{piasecki2024ai, prajod2026full}.

Transparency is widely regarded as a precondition for trustworthy AI, and journalism ethics frameworks have extended this principle to AI-assisted content production~\cite{koliska2022trust}. Current best practices typically involve textual labels ranging from brief tags (``this article was generated with AI assistance'') to detailed disclosure statements specifying the AI tools used, the nature of human oversight, and channels for reporting errors~\cite{prajod2026full, valenzuela2026effects, NordicAI}. These practices reflect journalistic guidelines about what constitutes responsible disclosure, but readers' perceptions are underexplored.
 
More recently, a shift toward interactive and progressive disclosure designs has been emerging. Kusters et al.~\cite{kusters2026more} designed and evaluated interactive visualization-based disclosure prototypes for AI-assisted news, including task-timelines and chatbot-style disclosures. Their work demonstrates that the design space for AI disclosure extends well beyond text labels. The participant-suggested designs I discuss below build on this direction.
 
%% ─────────────────────────────────────────────────────────────
\section{Position: Disclosure Is a Design Problem, Not a Compliance Checkbox}
%% ─────────────────────────────────────────────────────────────

My position rests on three empirical observations from~\cite{prajod2026full, prajod2026towards}, which I connect to a design argument.
 
\subsection{The Journalism-Reader Disconnect}
 
Journalistic principles hold that detailed AI disclosures (e.g., specifying human oversight, editorial accountability, and error-reporting channels) help readers make informed trust-based judgments. This assumption was directly tested in a controlled experiment with 34 participants. Detailed disclosures triggered what researchers call the \emph{transparency dilemma}: participants who read news with detailed disclosure information reported \emph{lower} trust than those who got simple one-line disclosures. One-line disclosures, in turn, did not reduce trust compared to no disclosure, but they carry a different cost. A brief label that acknowledges AI use without elaboration can trigger an information gap effect~\cite{loewenstein1994psychology}: readers know AI was involved but not how, prompting them to spend additional cognitive effort scanning the text for signs of AI involvement.
 
This is not an argument against transparency. It is evidence that \emph{what journalists think is good disclosure does not match what readers experience as trustworthy}. Some individual elements (e.g., knowing a human was involved) were valued in isolation, but the overall effect of the detailed disclosure package was negative. There is a genuine disconnect between the practitioner side and the user side of this interface.
 
This disconnect carries a further risk. If detailed disclosures become standard practice but readers routinely scroll past them or assume that the presence of a disclosure block means the outlet is being transparent without actually reading its contents, then these disclosures risk functioning as \textit{dark patterns}. That is, disclosures are technically present, practically invisible, and serving the outlet's compliance needs rather than the reader's informational needs. Until now, studies have not tested for dark patterns, but the interview insights from my study suggests the conditions for deceptive patterns are already in place.
 
\subsection{What Readers Actually Want: Agency and Design}
 
When participants were asked about their \emph{ideal} AI use disclosures in news, they did not ask for less information. They described fundamentally different disclosure \emph{designs} that give them agency over when and how much they engage with transparency information:
 
\begin{enumerate}
    \item \textbf{Highlight-for-glancing:} Highlighting key steps in which AI was involved while maintaining higher level of detail for a thorough inspection if needed. This design allows for a quick visual scan to grasp AI involvement without interrupting the reading flow.
    \item \textbf{Info button:} A brief disclosures with a small interactive element (e.g., an ``i'' icon) that readers can click or hover over to access detailed disclosure information. \emph{The readers choose when} they read detailed information (e.g., based on the news type, topic, or their own interest) rather than having it imposed upfront.
    \item \textbf{Outlet-level disclosure:} Transparency at the news organization level (``this outlet uses AI in the following ways...'') rather than per-article labels, reducing repetitive disclosure fatigue. Additionally, the readers can read this statement whenever they have the need to understand the outlet's AI use policy. This design also is in line with existing work that outlet's brand reputation matters when it comes to trust with AI use~\cite{liu2019machine}.
    \item \textbf{Proportional AI-ratio visualization:} A visual representation, such as partially colored page (e.g., 30\% of the page highlighted for 30\% AI contribution) or a percentage bar, showing the proportion of AI involvement. This visualization along with a brief statement such as ``AI was used in the final editorial layer'' would help visualize how thick the AI layer is.
    \item \textbf{Visual trust stamp:} A recognizable seal or badge indicating responsible AI use, analogous to organic food labels or verified account badges.
    \item \textbf{``No AI used'' label:} An explicit statement that no AI tools were used, reframing disclosure as a signal that works in both directions.
\end{enumerate}
 
\begin{figure*}[p!]
    \centering
    %% TODO: Replace with actual mockup figure
    \includegraphics[width=\linewidth]{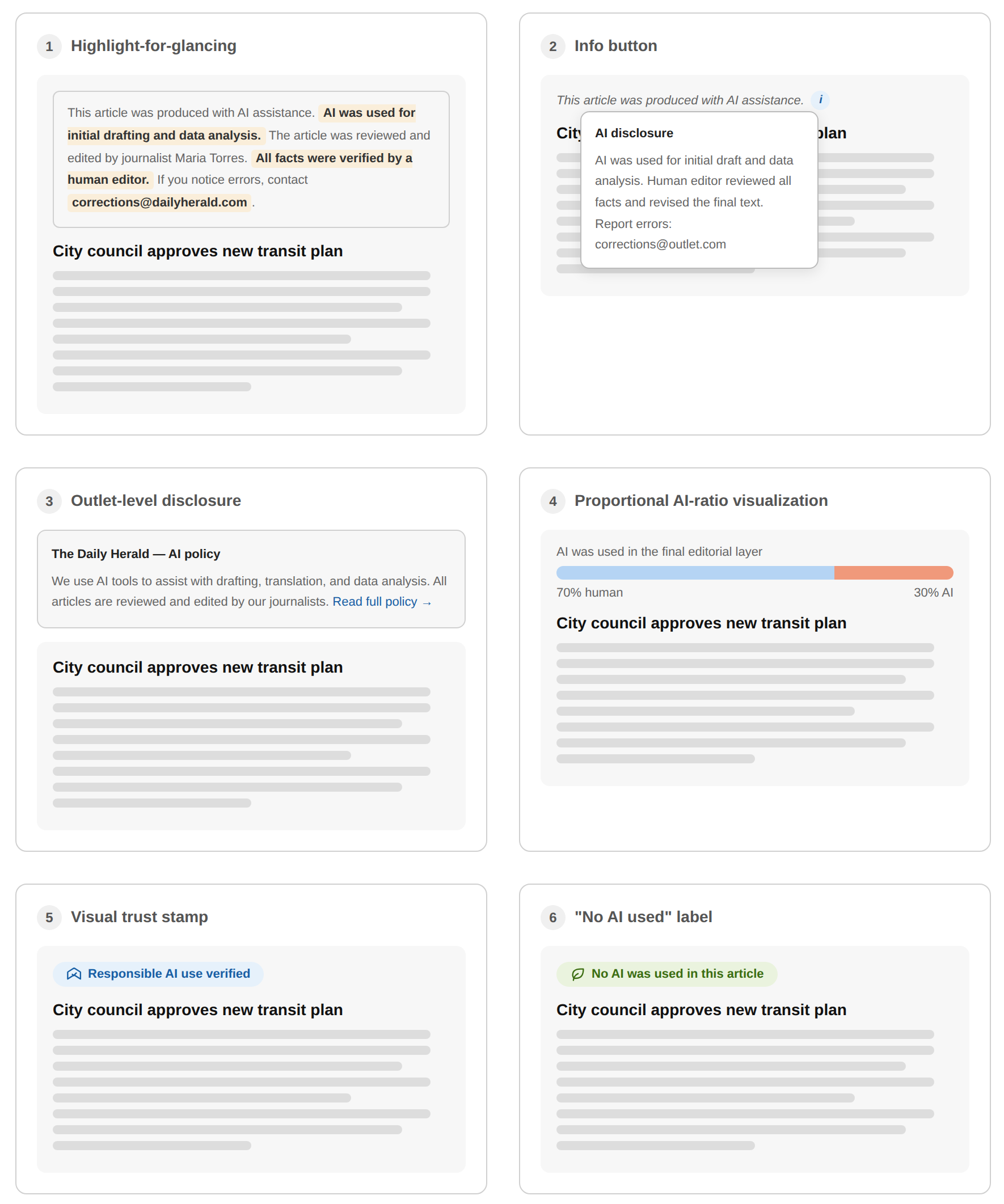}
    \caption{Mockups of six disclosure designs suggested by participants in~\cite{prajod2026full}. Info icon, highlight-for-glancing, and outlet-level disclosures give readers agency over when and how much disclosure they engage with. Proportional visualization convey AI involvement as a spectrum. Visual stamp and ``no AI'' labels reframe transparency from a warning to a broader trust signal.}
    \label{fig:mockups}
\end{figure*}

Figure~\ref{fig:mockups} visualizes these six concepts as mockups. %% TODO: Create and include figure

The recurring theme is \emph{user agency}. Some version info icon was suggested by multiple participants. Many of these designs let readers decide when disclosure matters to them rather than forcing a one-size-fits-all block of text. Such detail-on-demand designs resolve the core tension: the information \emph{exists} for those who want it, without triggering the transparency dilemma for those who do not.
 
\subsection{Readers Want Regulation Too}
 
Despite the AI Act's exemption for text content, 71\% of participants (24 of 34) stated that AI disclosure in news should be governmentally regulated. Readers do not share the regulation's assumption that human oversight makes disclosure unnecessary, they want mandatory disclosure for text too. But as the transparency dilemma shows, mandating disclosure is meaningless if the disclosure itself backfires. The challenge is not just \emph{whether} to require disclosure, but \emph{what form} that disclosure should take.

%% ─────────────────────────────────────────────────────────────
\section{Implications and Discussion Points}
%% ─────────────────────────────────────────────────────────────
 
The transparency dilemma in journalism raises uncomfortable questions for our own field. I bring two to the workshop:
 
\textbf{Are our own disclosure norms any better?} ACM now requires AI use statements in papers. Conference reviewers read them; readers encounter them. But has anyone tested whether these statements actually help reviewers or readers make better judgments about the work or infer the extent of AI involvement? They follow the same logic as the journalistic disclosures, i.e., we write them based on what \emph{we} think is responsible, not based on evidence of how they are received. If detailed AI disclosures backfire in news, the same could be true for AI use statements in research. We may be institutionalizing a transparency dilemma or even dark patterns in our own publishing norms.
 
\textbf{Do we design disclosures for ourselves or for our users?} A core finding from my work is that journalists' professional instincts about good disclosure do not match what readers actually need. This is a pattern worth examining beyond journalism. When HCI practitioners build AI-augmented tools, how do we decide what to disclose and how? Do we test those decisions against the people who use our outputs, or do we rely on professional instincts?

%% ─────────────────────────────────────────────────────────────
%% AI Use Statement (required)
%% ─────────────────────────────────────────────────────────────
\begin{acks}
  \textbf{AI Use Statement:} I used Claude Opus 4.6 (Anthropic) to assist with structuring and drafting this position paper. Figure~\ref{fig:mockups} was created using Claude based on my descriptions of each disclosure design, with the layout, placement, and visual choices iteratively refined through multiple rounds of feedback. The core argument, empirical findings, and design concepts are drawn from my own research. All claims, framing, and analysis are my own. I reviewed and revised all AI-generated content.

  This work was supported by the AI, Media \& Democracy Lab (Dutch Research Council project number: NWA.1332.20.009).
\end{acks}

%% ─────────────────────────────────────────────────────────────
\bibliographystyle{ACM-Reference-Format}
\bibliography{references}
%% ─────────────────────────────────────────────────────────────

\end{document}